\documentclass[%
 reprint,
superscriptaddress,
 amsmath,amssymb,
 aps,
]{revtex4-1}
\usepackage{ragged2e} 
\usepackage{booktabs}
\usepackage{multirow}
\usepackage{rotating}
\usepackage{subcaption}

\usepackage{float}
\usepackage{caption}
\usepackage{bm} 
\usepackage{graphicx}
\usepackage{dcolumn}
\usepackage{bm}
\usepackage{hyperref}

\begin{document}

\preprint{APS/123-QED}

\title{Detecting Black hole surrounded by perfect fluid dark matter in Kalb-Ramond fields using quasinormal modes}
\author{Zongyuan Qin}
\affiliation{%
 College of Physics,Guizhou University,Guiyang,550025,China
}%

\author{Taiyang Zhang}
\affiliation{Department of Physics, Guizhou University, Guiyang, 550025, China}

\author{Qian Feng}
\affiliation{Department of Physics, Guizhou University, Guiyang, 550025, China}

\author{Zheng-Wen Long}%
\email{zwlong@gzu.edu.cn (corresponding author)}
\affiliation{%
 College of Physics,Guizhou University,Guiyang,550025,China
}%


\begin{abstract}
	
This paper investigates the characteristics of quasinormal modes (QNMs) of static, spherically symmetric black holes under the combined influence of spontaneous Lorentz symmetry breaking (LSB) induced by the Kalb-Ramond (KR) field and perfect fluid dark matter (PFDM). Using M87* shadow data from the Event Horizon Telescope (EHT), we constrain the LSB factor $\tau$ and PFDM parameter $\zeta$ at 1$\sigma$ confidence. By combining the sixth-order WKB approximation method with time-domain numerical integration, we systematically compute the complex frequency spectrum of QNMs for black holes in this spacetime background. The numerical results reveal an intriguing conclusion: as the LSB factor $\tau$ or the PFDM parameter $\zeta$ increases, both the real part and the absolute value of the imaginary part of the QNMs frequencies exhibit a monotonic increase, demonstrating a unique ”stiffening” effect. This characteristic stands in stark contrast to the decreasing trend of QNMs frequencies observed in models that consider only traditional dark matter, revealing the critical influence of the coupling between the KR field and PFDM on the dynamic evolution of black holes. This study not only enriches and deepens the understanding of black hole perturbation theory within the framework of modified gravity but also, by identifying the distinctive spectral features of QNMs, offers the potential to distinguish whether the KR field and dark matter are coupled in future observations. Thus, it provides a theoretical foundation for testing mechanisms of spacetime symmetry breaking beyond the standard model and for exploring the nature of dark matter.
\end{abstract}

\maketitle


\section{Introduction}

In the fields of cosmology and high-energy astrophysics, black holes are compact objects predicted by general relativity that possess extreme spacetime curvature and strong gravitational fields. Their unique physical properties render them central objects of study for testing general relativity, exploring the nature of gravity, and understanding the laws of cosmic evolution \cite{Konoplya:2011qq,Virbhadra:1999nm,Gralla:2019xty,Zhao:2024lts,Liang:2024geh,Bian:2025ifp}. The study of black holes has shifted in recent times from theoretical speculation to a new phase of observational verification. The Laser Interferometer Gravitational-Wave Observatory (LIGO) and the Virgo collaboration have successfully detected gravitational wave signals generated by the merger of binary black holes\cite{LIGOScientific:2016aoc}, and the Event Horizon Telescope (EHT) has achieved direct imaging of the shadows of the supermassive black holes M87* and Sgr A*\cite{EventHorizonTelescope:2019dse,EventHorizonTelescope:2019uob,EventHorizonTelescope:2019jan,EventHorizonTelescope:2019ths,EventHorizonTelescope:2019pgp,EventHorizonTelescope:2019ggy}. These landmark observational achievements not only provide conclusive evidence for the existence of black holes but also inaugurate a new epoch of gravitational wave astronomy and multi-messenger astronomy. According to the black hole no-hair theorem\cite{Johannsen:2013szh,Kobayashi:2025evr,Yazadjiev:2025ezx}, the quasinormal modes (QNMs) spectrum of a black hole depends solely on its fundamental parameters such as mass, angular momentum, and charge, and is independent of the initial conditions of the perturbation. Consequently, this core physical characteristic makes the QNMs spectrum an ideal probe for precisely measuring the physical parameters of black holes and verifying models of black hole spacetime geometry. It offers a viable path to overcome the limitations of direct observation and to delve into the stability, spacetime symmetries, and dynamical evolution laws of black holes. Recently, the LVK collaboration detected the GW250114 event\cite{Akyuz:2025seg}, featuring a network matched-filter signal-to-noise ratio of nearly 80—marking it as the most powerful binary black hole merger signal ever recorded. Such a high signal-to-noise ratio enables us to probe the post-merger ringdown phase with unprecedented precision and extract the QNMs information of the remnant black hole.

~~Lorentz symmetry, a theoretical cornerstone of both general relativity and the Standard Model, has long been regarded as a fundamental pillar of modern physics. It not only underpins the spacetime structure of Einstein's field equations and special relativity but also shapes the formulation of fundamental interactions in particle physics, ensuring that physical laws remain invariant across all inertial frames.  Although this symmetry has been verified across a wide range of energy scales—from terrestrial laboratories to high-energy colliders and astrophysical observations—it may be broken in cutting-edge theories that seek to unify quantum mechanics and general relativity, such as string theory, non-commutative geometry, and certain quantum gravity models \cite{Kostelecky:1988zi,Alfaro:2001rb,Horava:2009uw,Carroll:2001ws,Jacobson:2000xp,Dubovsky:2004ud,Bengochea:2008gz,Cohen:2006ky}. This potential breaking provides a crucial avenue for investigating extensions to the Standard Model. The breaking of Lorentz symmetry can generally be categorized into two forms: explicit and spontaneous, which are fundamentally distinct in their physical mechanisms and theoretical foundations. Explicit breaking refers to the direct introduction of terms into the theoretical framework that violate invariance under Lorentz transformations. This renders the Lagrangian itself non-invariant under Lorentz transformations, leading to physical laws that are no longer unchanged by such transformations\cite{Nilsson:2022mzq,Lafkih:2024qva,Bluhm:2019ato,Allahyari:2025sbt}. In most modern physical frameworks, this is considered an unnatural assumption. In contrast, spontaneous Lorentz symmetry breaking maintains that the Lagrangian density formally retains full Lorentz covariance\cite{Ahmed:2026qmh}. However, after specific tensor fields acquire non-zero vacuum expectation values\cite{Kostelecky:1989jw,Kostelecky:1989jp,Bailey:2006fd,Bluhm:2008yt}(VEV), the ground state (vacuum state) no longer preserves the original symmetry. This spontaneous breaking mechanism, analogous to the Higgs mechanism\cite{Albareti:2016cvx}, can be naturally integrated into an effective field theory framework while preserving the theory's calculability and self-consistency under perturbative expansion.

~~Besides the intrinsic physical parameters of the black hole itself, its surrounding astrophysical environment—particularly the presence and spatial distribution of dark matter (DM) \cite{Ferrer:2017xwm,Nampalliwar:2021tyz,Xu:2021dkv,Turner:1984nf}—can also significantly alter the spacetime geometry and influence the dynamical behavior of matter and light within it. Research indicates that dark matter constitutes the dominant component of galactic mass\cite{Persic:1995ru}, suggesting that stars and interstellar gas contribute only a small fraction of a galaxy's total mass. In the central regions of galaxies, where the gravitational potential well is the deepest, the dark matter density rises sharply; Consequently, the supermassive black hole located in the galactic center is likely surrounded by a high-density dark matter halo. Therefore, investigating the nature of dark matter and its coupling mechanism with compact objects has become one of the frontier topics in modern astrophysics and cosmology. Among numerous dark matter theoretical models\cite{Hu:2000ke,Bode:2000gq,Potapov:2016obe,Spergel:1999mh}, PFDM, as a phenomenological description, is widely used to describe the matter distribution around black holes. This is because it can well explain the flattened features of the rotation curves of most spiral galaxies on large scales without introducing complex core-halo structures or non-standard dynamics. By incorporating PFDM into the Einstein field equations, one can obtain physically self-consistent dark matter-black hole spacetime solutions. This allows for a quantitative study of dark matter's impact on key observables\cite{Shodikulov:2025xax,Jumaniyozov:2025dyy} such as the black hole event horizon, black hole shadow, photon spheres, and QNMs, hence providing a useful theoretical avenue for investigating how dark matter and black holes interact.

~~Black holes in the real universe are often not isolated but are immersed in a complex material environment. Supermassive black holes at the centers of galaxies may not only be influenced by the LSB effect induced by the KR field but also inevitably interact with the surrounding DM halo. Although numerous studies have separately explored the impact of LSB\cite{Deng:2025atg,Al-Badawi:2023xig,Baruah:2023rhd}or DM\cite{Tan:2025usr,Liang:2024geh,Liu:2024xcd,Xavier:2023exm,Ma:2024oqe}on the spacetime geometry of black holes and their QNMss spectra, the dynamical behavior of black holes under the combined effects of both has, until now, lacked systematic investigation. Against this background, this paper constructs a static, spherically symmetric black hole solution involving both KR field-induced LSB and PFDM. We constrain the theoretical parameters using observational data of the M87* black hole shadow from the EHT\cite{EventHorizonTelescope:2022xqj,EventHorizonTelescope:2021dqv}. On this basis, we systematically calculate the QNMs spectra under scalar field, electromagnetic field, and axial gravitational field perturbations. By analyzing the modulation effects of the parameters on the real parts and imaginary parts of the frequencies, we reveal a distinctive "stiffening" effect arising from their coupling. Comparing this with traditional DM models, we further discuss its potential implications for future gravitational wave detections and black hole parameter inversion.

~~This paper is structured in the following manner: Sect.\ref{sec:2} derives the static spherically symmetric black hole metric within the KR-PFDM framework and analyzes its geometric properties. EHT observations of M87* are used to constrain the theoretical parameters. Within linear perturbation theory, wave equations for scalar, electromagnetic, and axial gravitational fields in this background are derived. Sect.\ref{sec:3} introduces the numerical methods adopted: the sixth-order WKB approximation and the time-domain integration method. Sect.\ref{sec:4} systematically investigates the influence of parameter variations on the QNMs spectra and analyzes the underlying physical mechanisms. Sect.\ref{sec:5} summarizes the findings and provides an outlook for subsequent work.

\section{Dynamic equations in the context of KR-PFDM spacetime background}
\label{sec:2}

In this paper, we consider a static, spherically symmetric black hole solution that is modified due to the coupling of two physical parameters: one is the PFDM distributed around the black hole, and the other is the background KR field. The KR field, originating from low-energy string theory, is a second-rank antisymmetric tensor field, of which non-zero VEV induces spontaneous LSB. The non-minimal coupling of this field with gravity modifies the spacetime geometry, resulting in deviations from the traditional black hole solutions derived in general relativity. Meanwhile, the presence of PFDM also affects the asymptotic behavior of the metric function. The spacetime obtained by combining these two effects is referred to as the KR-PFDM background spacetime, which reflects the geometric structure of spacetime under the coupling of the KR field and PFDM.~\cite{Lessa:2019bgi}

\begin{equation}
	\begin{split}
		S = \int d^4 x \sqrt{-g} \Biggl[  \frac{1}{2\kappa} (R - 2\Lambda + \epsilon B^{\mu\lambda}B^{\nu}_{\lambda}R_{\mu\nu})   \\
		- \frac{1}{12} H_{\lambda\mu\nu}H^{\lambda\mu\nu}- V(B_{\mu\nu}B^{\mu\nu} \pm b^2) + \mathcal{L}_{dm} \Biggr]
		\label{eq:1}
	\end{split}
\end{equation}

Here,$\kappa=8\pi G$  , where $G$ is Newton’s gravitational constant, $\Lambda$  is the cosmological constant, $\epsilon$  is the coupling constant between the KR field and gravity, and  $b^2$  is a real positive constant that characterizes the energy scale of spontaneous LSB. The KR field strength is expressed as $H_{\lambda\mu\nu} \equiv \partial_{[\lambda} B_{\mu\nu]}$, namely the antisymmetrized partial derivative. In~Eq.\eqref{eq:1}, $\mathcal{L}_{dm}$  represents the Lagrangian density of PFDM, and the self-interaction potential $V(B_{\mu\nu}B^{\mu\nu} \pm b^2)$  ensures spontaneous LSB, which gives rise to a nonvanishing VEV $\langle B_{\mu\nu} \rangle = b_{\mu\nu}$  for the KR field and satisfies the constant norm condition $b^{\mu\nu} b_{\mu\nu} = \mp b^2$, thus leading to a vanishing field strength of the KR field. The field equations follow from varying Eq.\eqref{eq:1} with respect to the metric $g^{\mu\nu}$:
\begin{equation}
R_{\mu\nu} - \frac{1}{2}g_{\mu\nu}R + \Lambda g_{\mu\nu} = \kappa \left( T_{\mu\nu}^{KR} + T_{\mu\nu}^{DM} \right)
\label{eq:2}
\end{equation}
Where~$T_{\mu\nu}^{DM}$  is the energy-momentum tensor of PFDM, given by:
\begin{equation}
	\begin{split}
		T_{\mu}^{\nu(DM)} &= \text{diag}\left[\frac{\zeta}{8\pi r^3}, \frac{\zeta}{8\pi r^3}, -\frac{\zeta}{16\pi r^3}, -\frac{\zeta}{16\pi r^3}\right]
	\end{split}
	\label{eq:3}
\end{equation}
\begin{equation}
	\begin{split}
		\kappa T_{\mu\nu}^{KR} =& \frac{1}{2} H_{\mu\alpha\beta} H_{\nu}^{\alpha\beta} - \frac{1}{12} g_{\mu\nu} H^{\alpha\beta\rho} H_{\alpha\beta\rho} \\
		& + 2V'(X) B_{\alpha\mu} B_{\nu}^{\alpha} - g_{\mu\nu} V(X) \\
		& + \epsilon \Biggl[ \frac{1}{2} g_{\mu\nu} B^{\alpha\gamma} B^{\beta}_{\gamma} R_{\alpha\beta} - B^{\alpha}_{\mu} B^{\beta}_{\nu} R_{\alpha\beta} \\
		& \qquad - B^{\alpha\beta} B_{\nu\beta} R_{\mu\alpha} - B^{\alpha\beta} B_{\mu\beta} R_{\nu\alpha} \\
		& \qquad + \frac{1}{2} \nabla_{\alpha} \nabla_{\mu} (B^{\alpha\beta} B_{\nu\beta}) + \frac{1}{2} \nabla_{\alpha} \nabla_{\mu} (B^{\alpha\beta} B_{\mu\beta}) \\
		& \qquad - \frac{1}{2} \nabla^{\alpha} \nabla_{\alpha} (B_{\mu}^{\gamma} B_{\nu\gamma}) - \frac{1}{2} g_{\mu\nu} \nabla_{\alpha} \nabla_{\beta} (B^{\alpha\gamma} B_{\gamma}^{\beta})\Biggr]
	\end{split}
	\label{eq:4}
\end{equation}
~~Here, $\zeta$ is the PFDM parameter, with the prime $'$ indicating the derivative taken with respect to the argument of the relevant function. The Bianchi identities ensure that the combined tensor $T_{\mu\nu}^{KR} + T_{\mu\nu}^{DM}$  is conserved. In this paper, we take the cosmological constant $\Lambda = 0$ .After obtaining the static, spherically symmetric black hole solution to the field equations, we arrive at the following metric:
\begin{equation}
	\begin{split}
		ds^2 &\equiv g_{\mu\nu} dx^\mu dx^\nu   \\
		&=-a(r) dt^2 + b(r) dr^2 + r^2 d\theta^2 + r^2\sin^2\theta d\phi^2
	\end{split}
	\label{eq:5}
\end{equation}
Assuming that $a(r) \equiv 1/b(r) \equiv f(r)$, and by employing the field equations Eq.\eqref{eq:5}, one gets the differential equation:
\begin{equation}
	  f''(r) + \frac{2}{r} f'(r) + \frac{\zeta}{2r^3} = 0
\label{eq:6}
\end{equation}
\begin{equation}
	f''(r) + \left(\frac{1}{r} + \frac{1}{r\tau}\right)f'(r) - \frac{1-\gamma}{\gamma r^2}f(r) = 0 ,
\label{eq:7}
\end{equation}
Combining Eq. \eqref{eq:6} and Eq.\eqref{eq:7}, we obtain:
\begin{equation}
rf'(r) + f(r) - \frac{1}{1-\tau} - \frac{\zeta}{r(1-\tau)} = 0,
\label{eq:8}
\end{equation}
Solving this equation yields:
\begin{equation}
f(r) = \frac{1}{1 - \tau} + \frac{\zeta}{r(1 - \tau)} \ln r + \frac{c}{r},
\label{eq:9}
\end{equation}
~~To determine the integration constant $c$ , we require that the obtained black hole solution reduces to the Schwarzschild black hole solution under appropriate physical conditions. First, when both the LSB factor $\tau$  and the PFDM parameter $\zeta$  are zero, the metric should degenerate to the Schwarzschild solution, i.e., $f(r) = 1 - \frac{2M}{r}$ .Second, when only $\tau=0$ , the solution reduces to the Schwarzschild black hole surrounded purely by PFDM, whose metric form has been given in previous studies, including a logarithmic term $\ln\left( \frac{r}{|\zeta|} \right)$\cite{Vachher:2024ldc}  . Meanwhile, considering dimensional consistency, to ensure the logarithmic term is physically meaningful, a characteristic length scale must be chosen to render the argument dimensionless. The parameter $\zeta$ precisely provides this scale. Based on these considerations, we take the constant $c$ as:$c = -2M - \frac{\zeta}{1-\tau} \ln\left(\left|\zeta\right|\right)$. Based on the above procedure, we obtain the explicit metric function $a(r) = \frac{1}{1 - \tau} - \frac{2M}{r} + \frac{\zeta}{r(1 - \tau)} \ln\left(\frac{r}{|\zeta|}\right)$~\cite{Jha:2025uie}, where $\tau = \frac{\varepsilon b^2}{2}$ denotes the LSB factor. This metric reduces to the Schwarzschild metric when $\tau \to 0$ and $\zeta \to 0$. Setting a(r) = 0 yields the horizon position equation:
\begin{equation}
	r_h = \zeta \cdot W(e^{-\frac{2M(\tau-1)}{\zeta}})
	\label{eq:10}
\end{equation}
where $W$ denotes the Lamber $W$ function.
\begin{figure*}[htbp]
	a)\includegraphics[width=8.1 cm]{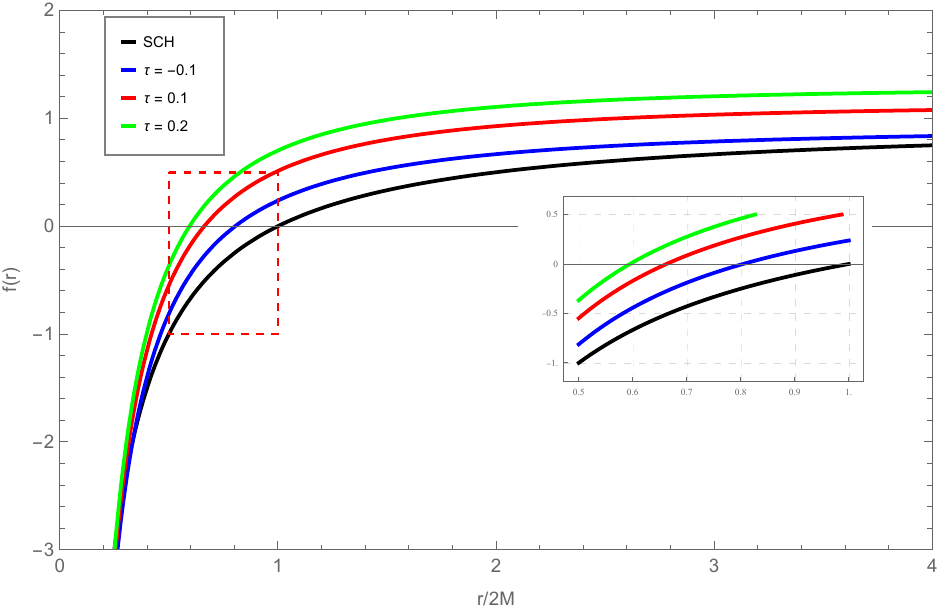}
	b)\includegraphics[width=8.1 cm]{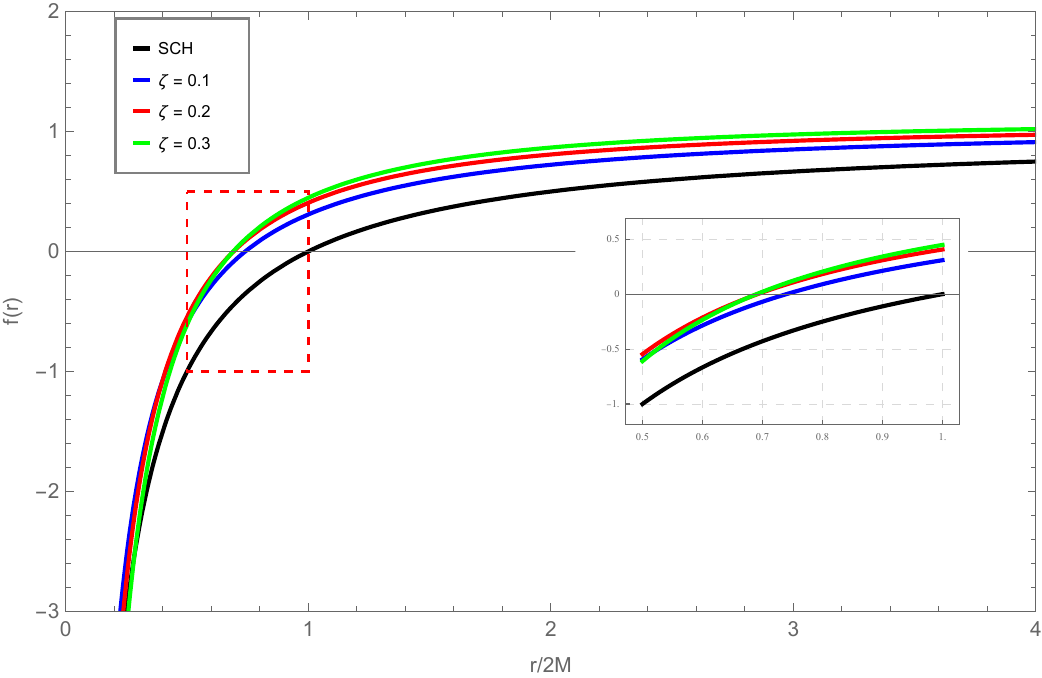}
	\caption{Shows the variation of the black hole metric function in the KR-PFDM spacetime background with parameters $\tau$ and $\zeta$. The Schwarzschild black hole metric is depicted by the black curve. Fixing $M = \frac{1}{2}$, we set $\zeta=0.3$ to investigate the metric function variation for different values of $\tau$ (left panel), and set $\tau = 0.06$ to research the metric function variation for different values of $\zeta$ (right panel).} \label{fig: 1}
\end{figure*}
\begin{figure*}[htbp]
	a)\includegraphics[width=8.1 cm]{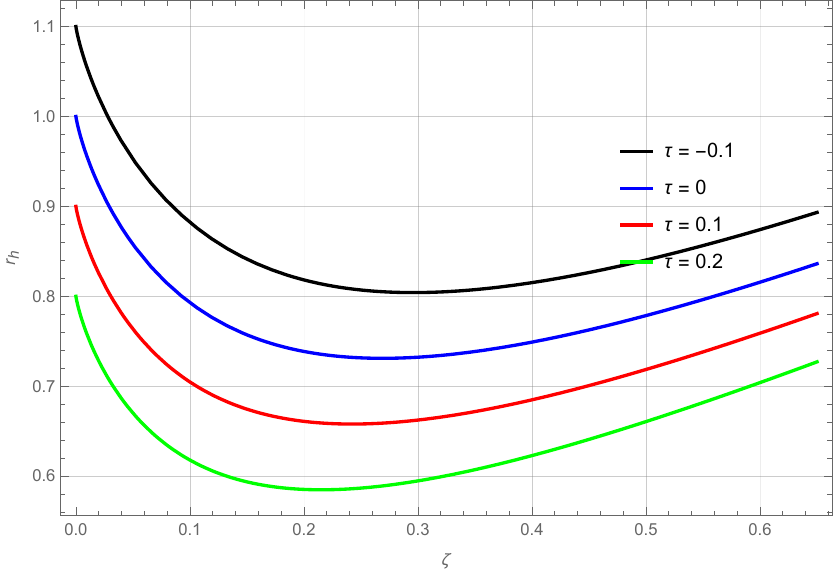}
	b)\includegraphics[width=8.1 cm]{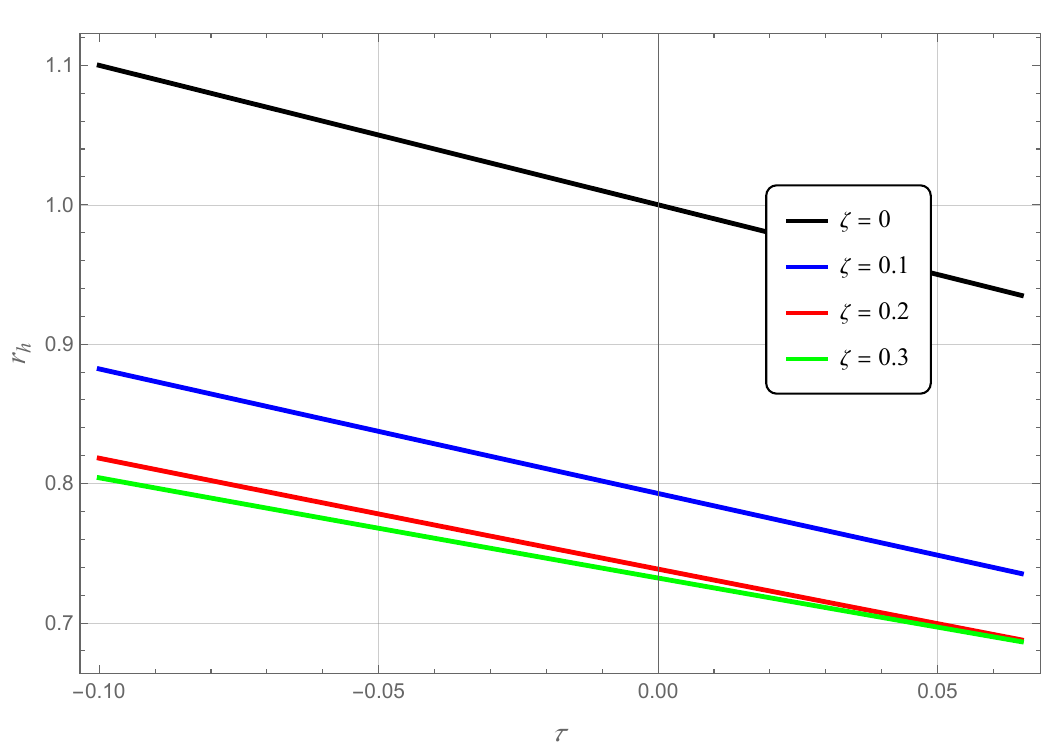}
	\caption{The variation of the event horizon as a function of parameters $\tau$ and $\zeta$ in the KR-PFDM spacetime background. For fixed $M = \frac{1}{2}$ , the variation of the event horizon with $\tau$ for different values of $\zeta$ is shown (right panel), and the variation with $\zeta$ for different values of $\tau$ is shown (left panel).} \label{fig: 2}
\end{figure*}

Fig.\ref{fig: 1} displays the variation of the black hole metric function $f(r)$ with respect to the theoretical parameters $\tau$ and $\zeta$ in the KR-PFDM spacetime background. The black curve in the figure always represents the classical Schwarzschild black hole metric, serving as the benchmark for theoretical comparison. We fix the mass parameter $M = 1/2$  and investigate the effects of the two key model parameters $\tau$ and $\zeta$ on the metric structure, respectively. As $\tau$ increases, the shape of the metric function curve near the horizon region changes significantly; not only does the position of the event horizon shift, but the slope of the function near the horizon also varies accordingly, which is directly related to the temperature and thermodynamic properties of the black hole. With variations in $\zeta$, the behavior of the metric function near the event horizon also undergoes notable changes. When $\zeta$ takes positive real values, the metric function curve shifts upward overall compared to the Schwarzschild case, and its intersection point with the $f(r)=0$ axis (i.e., the event horizon position $r_h$ ) shifts inward correspondingly, indicating that the black hole's horizon radius decreases as the parameter $\zeta$ increases. Furthermore, all curves approach different constant values in the asymptotic region, implying that the parameter $\zeta$ modifies the asymptotic geometry of spacetime, corresponding to the correction to the far-field gravitational potential induced by the coupling effect of the KR field and PFDM. Fig.\ref{fig: 2} systematically demonstrates how the event horizon radius of the black hole varies with the theoretical parameters $\tau$ and $\zeta$ , intuitively revealing the geometric effects introduced by the newly incorporated modifications. Synthesizing Fig.\ref{fig: 1} and Fig.\ref{fig: 2}, it is evident that the event horizon of the black hole predicted by this theoretical model is no longer static but is sensitively dependent on the parameters $\tau$ and $\zeta$.These results not only provide an intuitive "phase diagram" for understanding black hole geometry within modified gravity theories but also establish a crucial theoretical foundation for the subsequent exploration of QNMs.

\subsection{Scalar field and electromagnetic field perturbations}
In this section, we consider the Klein-Gordon equations for massless scalar and electromagnetic fields in the background of this metric.
\begin{equation}
	\frac{1}{\sqrt{g}} \partial_\mu \left( \sqrt{g} g^{\mu\nu} \partial_\nu \Phi \right) = 0
	\label{eq:11}
\end{equation}
\begin{equation}
	\frac{1}{\sqrt{-g}} \partial_\mu (F_{\rho\sigma}g^{\rho\nu}g^{\sigma\nu} \sqrt{-g}) = 0
	\label{eq:12}
\end{equation}
	~~where g denotes the determinant of the metric tensor, $g^{\mu\nu}$ is the inverse metric, $F_{\rho\sigma} = \partial_\rho A_\sigma - \partial_\sigma A_\rho$ represents the electromagnetic field tensor, and $A_\nu$  is the electromagnetic four-potential. Under a spherically symmetric metric, the evolution of the fields exhibits no dependence on rotational degrees of freedom. Therefore, we can ignore the complex variations on the angular coordinates $\theta$ and $\phi$ , and perform the following decomposition for the wavefunctions:
	\begin{equation}
		\Phi(t,r,\theta,\phi) = \frac{e^{-i\omega t}}{r} \Psi(r) Y_{l}^{m}(\theta,\phi)
		\label{eq:13}
	\end{equation}
	~~Here, $\omega$ is the frequency of the wavefunction, $\Psi$ denotes the radial wavefunction, l is the multipole angular quantum number, and the magnetic quantum number m is assigned a value of zero in this study. $Y_{l}^{m}$  represents the Hamiltonian spherical harmonics. The radial equation corresponding to Eq.\eqref{eq:11} can be expressed in the following form:
	\begin{equation}
		\frac{1}{\sin\theta} \partial_\theta (\sin\theta \partial_\theta P_{lm}) - \frac{m^2}{\sin^2\theta} P_{lm} = -l(l+1) P_{lm}
		\label{eq:14}
	\end{equation}
	~~For the purpose of simplifying Eq.\eqref{eq:11} and Eq.\eqref{eq:12}, the tortoise coordinate is adopted.
	\begin{equation}
		dr_* = dr/f(r)
		\label{eq:15}
	\end{equation}
	~~By introducing the tortoise coordinate, the evolution equations for the scalar and electromagnetic fields are recast as a Schrödinger-like equation:
	\begin{equation}
		\frac{d^2 \psi_s}{dr_*^2} + (\omega^2 - V(r)) \psi_s = 0
		\label{eq:16}
	\end{equation}
	~~The explicit form of the effective potential is as follows:
		\begin{equation}
			V(r) = \frac{f(r)}{r^2} \left[ l(l+1) + (1-s) r f'(r) \right]
		\label{eq:17}
	\end{equation}
~~where s is the spin parameter. 
	
~~The effective potential for the scalar field (s=0) takes the form:
	\begin{equation}
		\begin{split}
			V_s(r) =  \left[ \frac{1}{1-\tau} - \frac{2M}{r} + \frac{\zeta}{r(1-\tau)} \ln\left(\frac{r}{|\zeta|}\right) \right] \\
			 \times\left[ \frac{l(l+1)}{r^2} + \frac{2M}{r^3} + \frac{\zeta}{(1-\tau) r^3} \left(1 - \ln\frac{r}{|\zeta|}\right) \right]
		\end{split}
		\label{eq:18}
	\end{equation}
	~~The effective potential for the electromagnetic field (s=1) takes the form:
	\begin{equation}
		V_E(r) = \left[ \frac{1}{1-\tau} - \frac{2M}{r} + \frac{\zeta}{r(1-\tau)} \ln\left(\frac{r}{|\zeta|}\right) \right] \cdot \frac{l(l+1)}{r^2}
		\label{eq:19}
	\end{equation}
\subsection{Axial gravitational perturbations of the black hole in the KR-PFDM spacetime background}
~~Gravitational field perturbations, also known as metric perturbations, are an important tool for studying black hole.For axial gravitational perturbations, the metric structure consists of the background metric describing the stationary spacetime and the linear perturbation part induced by external disturbances. Investigating such axial perturbation modes in the KR-PFDM spacetime background will help to understand the structural stability of this metric and its response characteristics to gravitational wave signals. Therefore, in this context, the metric tensor can be approximately expressed as:
\begin{equation}
	g_{\mu\nu} = \bar{g}_{\mu\nu} + h_{\mu\nu}
	\label{eq:20}
\end{equation}
	~~In the KR-PFDM spacetime background, we express the metric as:
	\begin{equation}
		\bar{g}_{\mu\nu} = \text{diag}(-f(r), \frac{1}{f(r)}, r^2, r^2 \sin^2 \theta)
		\label{eq:21}
	\end{equation}
	~~In the KR-PFDM model, the function f(r) describes the geometric structure of this spacetime. The term $h_{\mu\nu}$ represents the linear perturbation, satisfying $\bar{g}_{\mu\nu} \gg h_{\mu\nu}$ . Based on its symmetry properties, this perturbation can be further decomposed into two types of modes: axial modes and polar modes.
	\begin{equation}
	h_{\mu\nu} = h_{\mu\nu}^{\text{axial}} + h_{\mu\nu}^{\text{polar}}
		\label{eq:22}
	\end{equation}
	~~This paper considers only axial gravitational perturbations for computational simplicity. For perturbations of this type, the Regge-Wheeler (RW) gauge \cite{Regge:1957td} is adopted, and the perturbed metric is expressed as:
	\begin{equation}
		h_{\mu\nu}^{\text{axial}} = \begin{pmatrix} 0 & 0 & 0 & h_0 \\ 0 & 0 & 0 & h_1 \\ 0 & 0 & 0 & 0 \\ h_0 & h_1 & 0 & 0 \end{pmatrix} \sin\theta \partial_\theta P_l(\cos\theta)
		\label{eq:23}
	\end{equation}
	~~Here, $P_l(\cos\theta)$ denotes the Legendre polynomial of order l , which describes the angular distribution of axial perturbation modes. For a static spherically symmetric black hole, the magnetic quantum number is taken as m=0, and $h_0$ and $h_1$ are functions solely of the time coordinate t and radial coordinate r.
	
	~~Under this metric, the Christoffel symbols are modified as follows:
	\begin{equation}
		\Gamma^{\mu}_{\nu\lambda} =\bar{\Gamma}^{\mu}_{\nu\lambda} + \delta\Gamma^{\mu}_{\nu\lambda}
		\label{eq:24}
	\end{equation}
	~~In perturbation analysis, $\bar{\Gamma}^{\mu}_{\nu\lambda}$ represents the Christoffel symbols associated with this metric, while the linear correction term induced by the perturbation $h_{\mu\nu}$ can be expressed as $\delta\Gamma^{\mu}_{\nu\lambda}$, whose explicit form is given below:
	\begin{equation}
		\delta\Gamma^{\mu}_{\nu\lambda} = \frac{1}{2} \bar{g}^{\mu\rho} (h_{\rho\lambda;\nu} + h_{\rho\nu;\lambda} - h_{\nu\lambda;\rho})
		\label{eq:25}
	\end{equation}
	~~thereby, we rewrite the Ricci tensor as:
	\begin{equation}
		R_{\mu\nu} = \bar{R}_{\mu\nu} + \delta R_{\mu\nu}
		\label{eq:26}
	\end{equation}
	~~Neglecting the coupling between matter fields\cite{Konoplya:2003ii}, the axial gravitational perturbation equation can be expressed as:
	\begin{equation}
		\delta R_{\mu\nu} = 0
		\label{eq:27}
	\end{equation}
	~~From Eq.\eqref{eq:27}, we obtain $\delta R_{\theta\phi} = 0$,$\delta R_{r\phi} = 0$. 
	Further derivation yields:
	\begin{equation}
			\dot{h}_0 = (f h_1)'
		\label{eq:28}
	\end{equation}
	\begin{equation}
		\frac{1}{f} \ddot{h}_1 - \dot{h}_0' + \frac{2}{r} \dot{h}_0 + \frac{l(l+1)-2}{r^2} f h_1 = 0
		\label{eq:29}
	\end{equation}
	~~The symbol · denotes the partial derivative with respect to t, the symbol ·· denotes the second-order partial derivative with respect to t, and the symbol ' denotes the partial derivative with respect to r. Introducing the variable $\Psi(t,r) = \frac{f(r)}{r} h_1(t,r)$ and the tortoise coordinate $dr_* = dr/f(r)$, we obtain the wave equation for axial gravitational perturbations:
	\begin{equation}
		\frac{\partial^2}{\partial t^2} \Psi(t,r) - \frac{\partial^2}{\partial r_*^2} \Psi(t,r) + V_G(r) \Psi(t,r) = 0
		\label{eq:30}
	\end{equation}
	~~Where $V_G(r)$ is the effective potential for gravitational perturbations, and its explicit expression is given by:
	\begin{equation}
		\begin{split}
			V_G(r) = & \left[ \frac{1}{1-\tau} - \frac{2M}{r} + \frac{\zeta}{r(1-\tau)} \ln\left( \frac{r}{|\zeta|} \right) \right] \\
			& \times \biggl[ \frac{l(l+1)}{r^2} + \frac{2\tau}{r^2(1-\tau)} - \frac{6M}{r^3} \\
			& \qquad + \frac{\zeta}{r^3(1-\tau)} \left( 3\ln\left( \frac{r}{|\zeta|} \right) - 1 \right) \biggr]
		\end{split}
		\label{eq:31}
	\end{equation}
	\begin{figure*}[htbp]
		a)\includegraphics[width=8.1 cm]{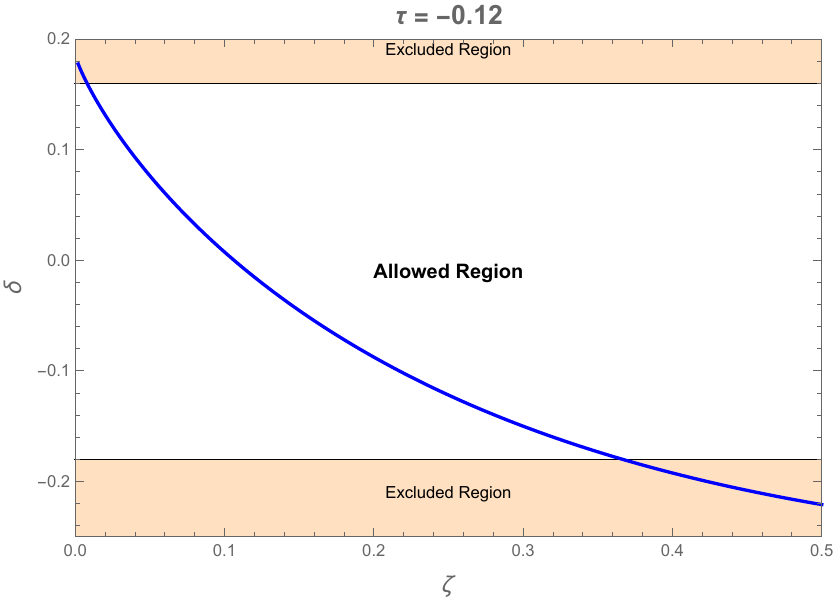}
		b)\includegraphics[width=8.1 cm]{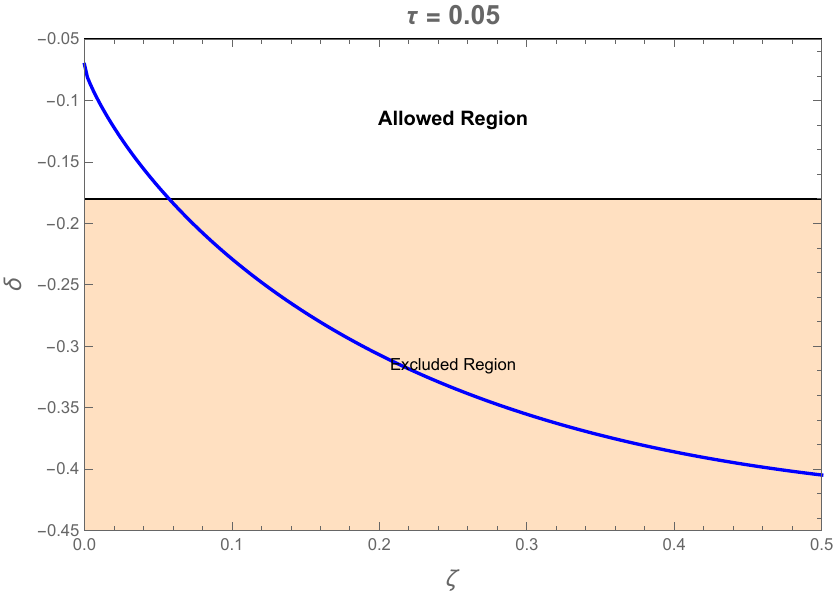}
		\caption{For the cases of $\tau=-0.12$ (left panel) and $\tau=0.05$ (right panel), this paper presents the distribution of the Schwarzschild deviation parameter $\sigma$  as a function of $\zeta$.The white regions in the figure indicate that within these parameter ranges, our black hole model is consistent with the EHT observations at the 1$\sigma$ confidence level.} \label{fig: 3}
	\end{figure*}
	~~Based on Fig.\ref{fig: 3} and RE.\cite{Jha:2025uie}, combined with the EHT observations of M87*, we constrain the permissible ranges of the LSB factor $\tau$ and the PFDM parameter $\zeta$ to $\tau \in (-0.1, 0.05)$ and $\zeta \in (0, 0.05)$ , respectively. Our study imposes stringent bounds on the parameters  $\tau$ and $\zeta$ , establishing our model as a viable candidate for supermassive black holes (SMBHs). All subsequent investigations in this paper will be conducted within this range.
	
	~~Based on the KR-PFDM theoretical framework, we preliminarily investigate the relevant properties of the effective potential in this spacetime. The morphological features of the effective potential are directly related to the QNMs spectra of the perturbed black hole, with its peak structure, width, and amplitude collectively influencing both the oscillation frequency and the damping rate. In this model, the LSB factor $\tau$ and the PFDM parameter $\zeta$ serve as key variables that jointly influence the spacetime geometry and its response behavior to external perturbations.
	
\begin{figure*}[htbp]
			\centering
			\includegraphics[width=0.3\textwidth]{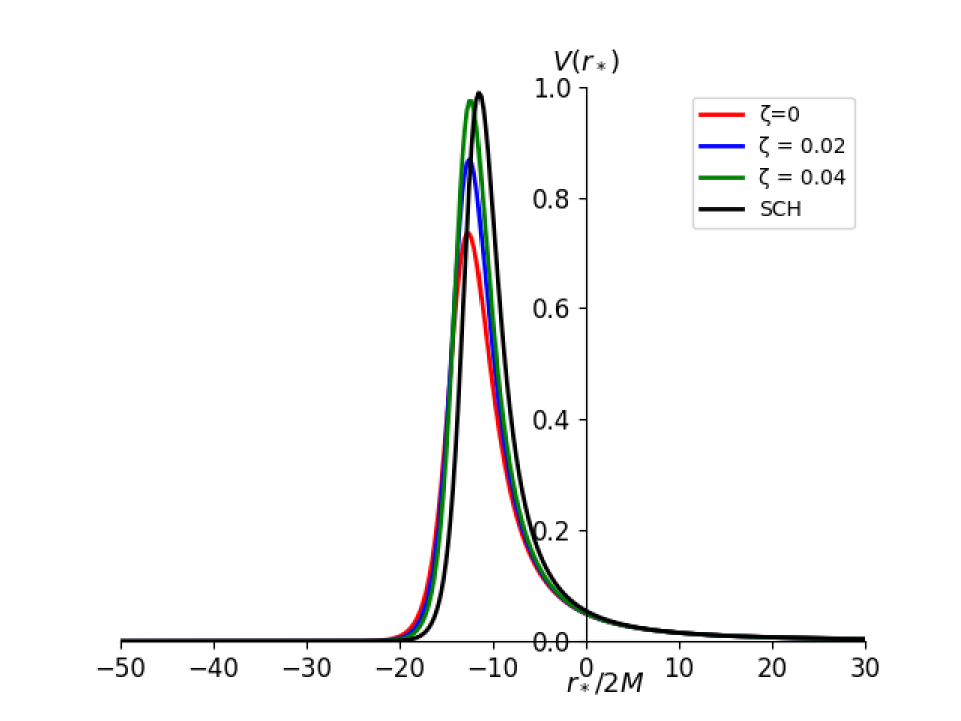} 
			\hfill
			\includegraphics[width=0.3\textwidth]{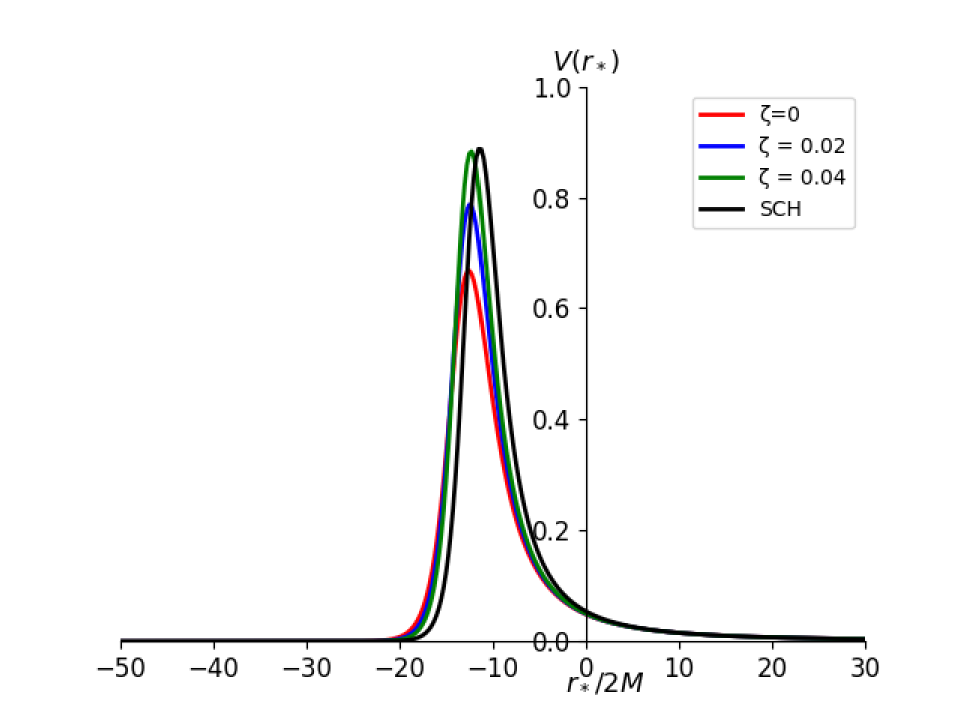} 
			\hfill
			\includegraphics[width=0.3\textwidth]{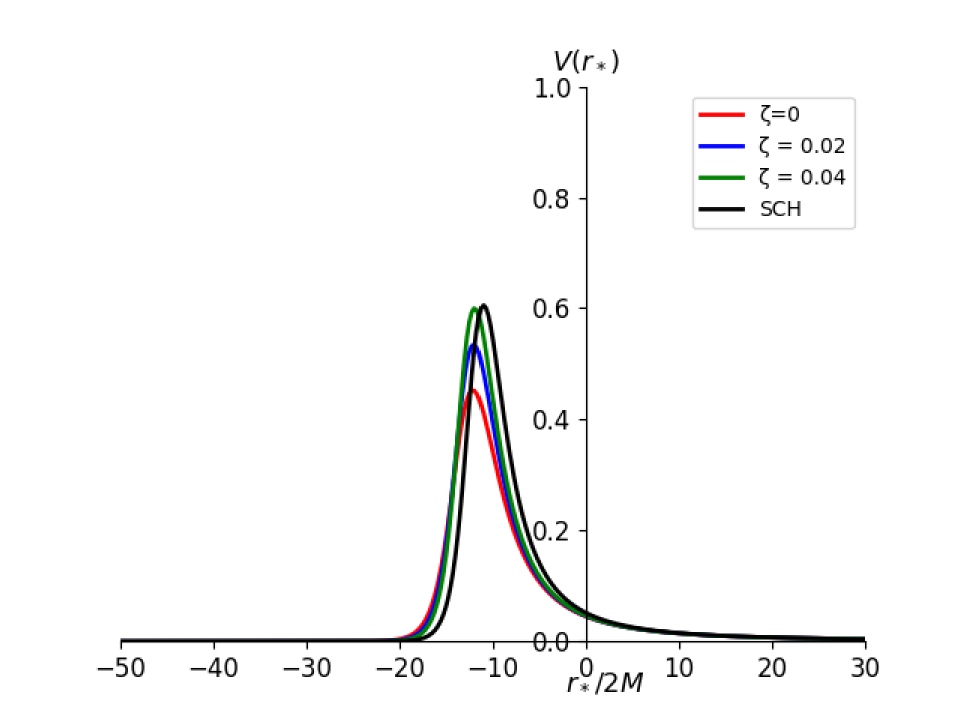} 
			\caption{Displayed is the variation of the effective potential with the tortoise coordinate under scalar field (left), electromagnetic field (middle), and gravitational field (right) perturbations for the black hole in the KR-PFDM spacetime. The parameters $M = 1/2$, $l=2$, and $\tau = -0.1$ are adopted, and the variation of the effective potential to different $\zeta$ values is analyzed.}
			\label{fig:4}
		\end{figure*}

\begin{figure*}[htbp]
			\centering
			\includegraphics[width=0.3\textwidth]{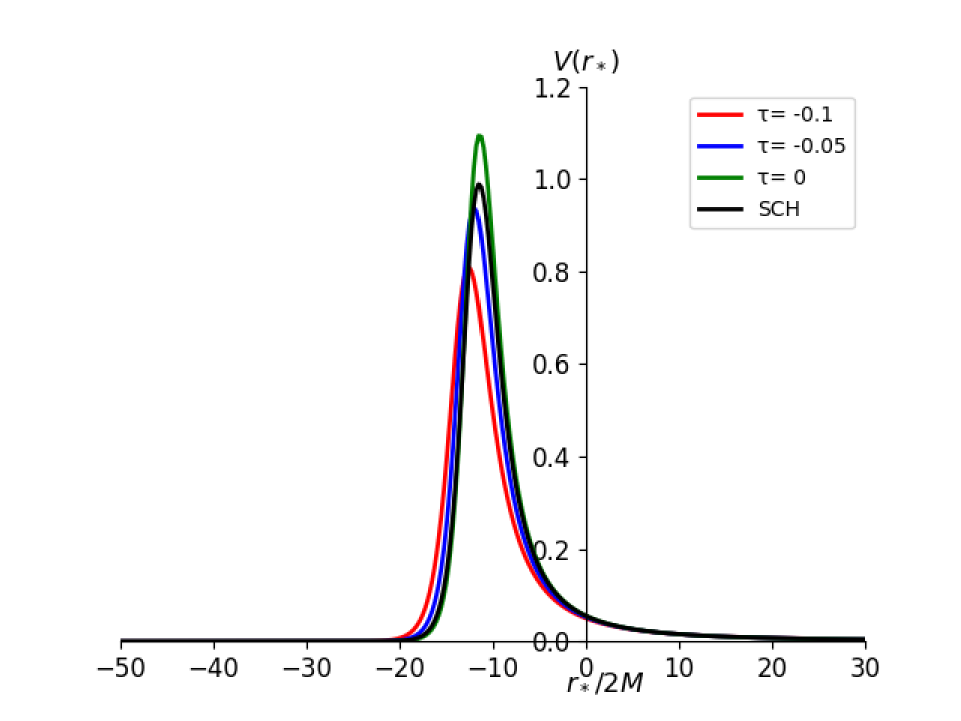} 
			\hfill
			\includegraphics[width=0.3\textwidth]{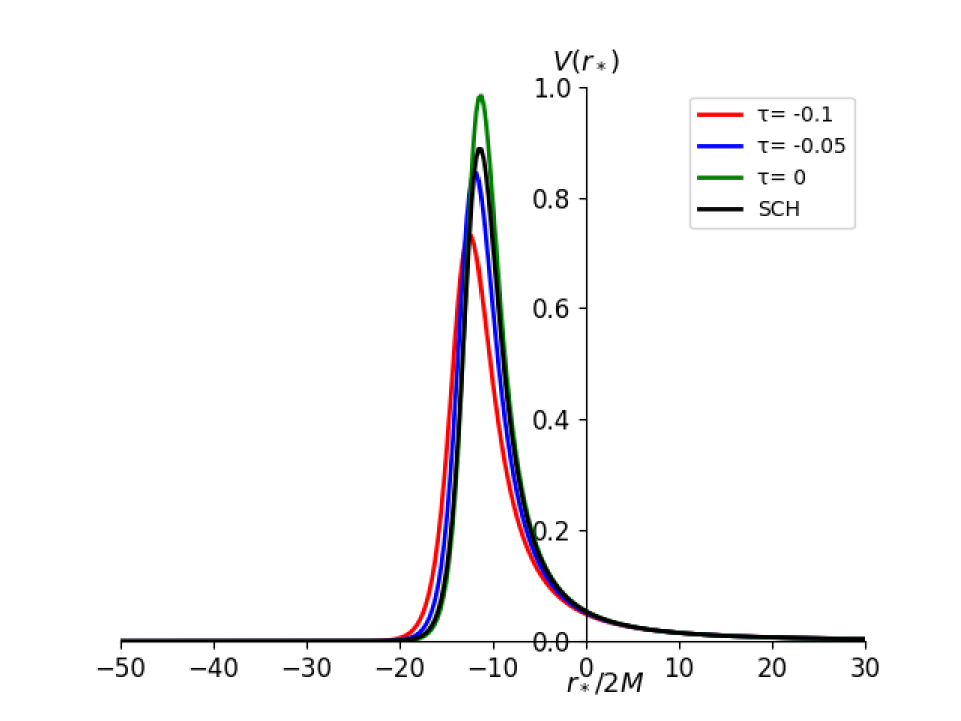} 
			\hfill
			\includegraphics[width=0.3\textwidth]{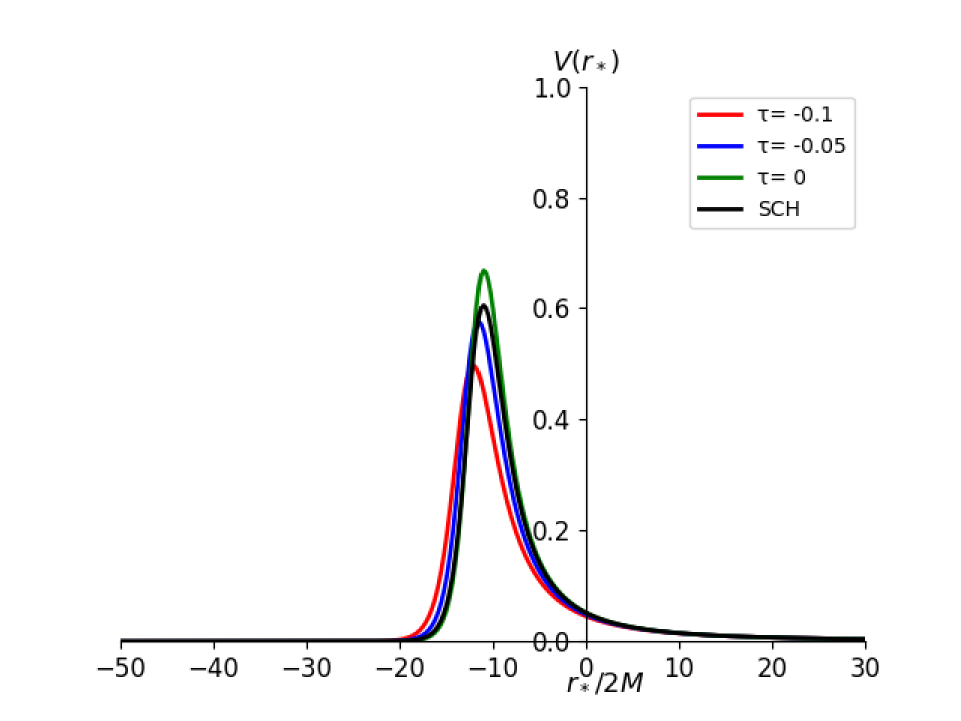} 
			\caption{Displayed is the variation of the effective potential with the tortoise coordinate under scalar field (left), electromagnetic field (middle), and gravitational field (right) perturbations for the black hole in the KR-PFDM spacetime. The parameters $M = 1/2$, $l=2$, and $\tau = -0.1$ are adopted, and the variation of the effective potential to different $\tau$ values is analyzed.}
			\label{fig:5}
		\end{figure*}

	~~Fig.\ref{fig:4} and Fig.\ref{fig:5} reveal the dependence of the effective potential distribution curves corresponding to different perturbation fields on the model parameters. As the PFDM parameter $\zeta$ increases, the peak of the effective potential exhibits a monotonically increasing trend, indicating that the introduction of this parameter enhances the height of the potential barrier to a certain extent. Meanwhile, it is noted that an increase in the LSB factor $\tau$ also leads to a monotonically increasing trend in the potential barrier height, suggesting that a stronger LSB effect similarly contributes to enhancing the spacetime's ability to confine external perturbations. In particular, for the parameter combination $\tau=0$, $\zeta=0.01$, the peak of the effective potential is slightly higher than that of the Schwarzschild black hole, implying that the spacetime structure under this parameter combination possesses a stronger capability to confine external perturbations. In Fig.\ref{fig:4} and Fig.\ref{fig:5}, the right, middle, and left panels correspond to the effective potential distribution curves for axial gravitational field perturbations (s=2), electromagnetic field perturbations (s=1), and scalar field perturbations (s=0), respectively. These figures clearly demonstrate the dependence of the potential barrier height on the spin parameter s : as the spin parameter s increases, the height of the potential barrier exhibits a significant decreasing trend. This pattern remains consistent across different values of $\tau$ and $\zeta$ , indicating that the spin effect plays a crucial role in modulating the perturbation behavior of black holes. Physically, a higher potential barrier generally implies that it is more difficult for perturbation waves to penetrate, and their energy dissipation rate is altered correspondingly, which directly affects the decay rate of the spacetime ringdown signal.
	
	~~Therefore, when the height of the potential barrier increases, the obstructive effect encountered by perturbation waves during propagation is enhanced, and the rate of energy dissipation correspondingly accelerates, which directly leads to a faster dynamic evolution of the spacetime structure near the black hole. This coupling mechanism between potential barrier evolution and black hole dynamics is the core physical factor that determines the spectral distribution characteristics and evolution laws of QNMs.

\section{Calculation methods for quasinormal modes}
\label{sec:3}

When a black hole is perturbed externally, it produces a series of characteristic eigen-oscillations known as quasinormal modes (QNMs)~\cite{Chandrasekhar:1984siy,Yang:2025hqk}. The complex frequencies corresponding to these oscillations precisely encode the response characteristics of the black hole spacetime to external perturbations. In gravitational wave detection, the ringdown phase represents the macroscopic manifestation of the superposition of these multi-order QNMs, thus serving as an important observational probe for revealing the geometric structure of black hole spacetimes and their surrounding environmental media. To systematically analyze the spectral characteristics and evolutionary behavior of QNMs in this spacetime background, this paper will adopt a combined strategy of the WKB (Wentzel–Kramers–Brillouin) approximation method and time-domain numerical integration to carry out the research.

\subsection{ WKB approximation method}
In the study of QNMs, the WKB approximation method has become a widely used 
semi-analytical approach due to its efficiency and simplicity in solving wave equations. This method was originally proposed by Schutz and Will~\cite{Schutz:1985km}, and later progressively developed to higher orders by scholars such as Iyer, Will, and Konoplya~\cite{Konoplya:2003ii,Iyer:1986nq,Iyer:1986np}. Matyjasek and Opala further advanced its precision to the 13th order by combining it with the Padé approximation\cite{Matyjasek:2017psv}. Owing to its efficiency and universality, the WKB approximation method is applicable to most black hole models~\cite{Konoplya:2003dd}. In practical applications, by separating the time variable through Fourier transformation and selecting appropriate spherical harmonic functions (scalar, vector, or tensor) to handle the angular components, the radial equation $\Psi(r)$ of the perturbed field can be transformed into a form amenable to analytical treatment. Based on the wave Eq.\eqref{eq:16} and Eq.\eqref{eq:30} derived earlier, we employ the WKB approximation method to compute the QNMs of black holes in the KR-PFDM spacetime background. To ensure the physical rationality of the equation solutions, we require that the wave function only has incoming waves at the horizon and only outgoing waves at spatial infinity, whose asymptotic behavior is given by the following formula:
\begin{equation}
\Psi(r_*) \sim e^{-i\omega r_*}, \quad r_* \to -\infty
	\label{eq:32}
\end{equation}
\begin{equation}
\Psi(r_*) \sim e^{i\omega r_*}, \quad r_* \to +\infty
	\label{eq:33}
\end{equation}
~~Here, $r_*$ denotes the tortoise coordinate, defined such that $\quad r_* \to -\infty$ as it approaches the event horizon, and $\quad r_* \to +\infty$ as it approaches spatial infinity. To solve for the QNMs frequencies using the sixth-order WKB approximation method, it is necessary to perform a piecewise expansion of the wavefunction under specific boundary conditions. This method first requires that, in the vicinity of the event horizon and at spatial infinity, the wavefunction satisfies the asymptotic behaviors of purely ingoing waves and purely outgoing waves, respectively. Secondly, the effective potential is expanded into a Taylor series at the peak of the potential barrier. By matching the WKB expansions of the near-field and far-field, the analytical expression of the QNMs eigenfrequencies can be derived. It should be noted that the applicability of this method presupposes that the effective potential exhibits a single-peak distribution and decays monotonically on both sides of the peak. Its calculation process can be further reduced to solving the characteristic equation of the following form:
\begin{equation}
\frac{i(\omega^2 - V_0)}{\sqrt{-2V_0''}} - \sum_{i=2}^{6} \Lambda_i = n + \frac{1}{2}
\label{eq:34}
\end{equation}

Here, $V_0$ represents the peak value of the effective potential, $V_0''$ denotes the second derivative of the effective potential with respect to the tortoise coordinate at the peak, $\Lambda_k(n)$ corresponds to the correction terms in the k-th order WKB method, and $r = r_*$ is the tortoise coordinate at which the effective potential for QNMs attains its maximum value. The calculation focus of this paper is on the frequency of the fundamental QNM, namely the case where the overtone number n=0 .

\subsection{Time-domain Method}

This paper employs the time-domain numerical method to investigate the dynamical behavior of the M87* black hole under gravitational perturbations. We first rewrite the wave Eq.\eqref{eq:30} for axial gravitational perturbations into a form expressed in terms of light-cone coordinates $u = t - r_*$ and $v = t + r_*$ .Through this coordinate transformation, the wave equation for axial gravitational perturbations originally established in the conventional coordinate system is converted into the light-cone coordinate system, thereby making it suitable for time-domain numerical calculations.

\begin{equation}
4\frac{\partial^2 \Psi}{\partial u \partial v} = -V(r)\Psi 
	\label{eq:35}
\end{equation}
~~In the numerical calculation, this paper adopts a discretization method to solve the partial differential equation. The computational domain is divided into uniform grids, where the grid spacing is denoted as $\Delta h$, and the grid nodes are $N = \left( u + \Delta h,\ v + \Delta h \right)$, $W = \left( u + \Delta h,\ v \right)$, $E = \left( u,\ v + \Delta h \right)$ and $S = \left( u,\ v \right)$ respectively. According to the numerical method proposed by Gundlach et al.~\cite{Gundlach:1993tp,Moderski:2001gt,Moderski:2005hf}, a second-order accurate discretization method is employed, whose expression is as follows:
\begin{equation}
	\begin{split}
		\Psi(N) = & \Psi(W) + \Psi(E) - \Psi(S)  \\
		& -(\Delta h)^2 \times \frac{V(W)\Psi(W) + V(E)\Psi(E)}{8} \\
		&+ o\bigl((\Delta h)^4\bigr)
	\end{split}
	\label{eq:36}
\end{equation}
~~The advantage of this discretization method lies in its combination of second-order accuracy and relatively high computational efficiency. The error term introduced by discretization is  , which primarily originates from the treatment of the mixed derivative term. According to the difference Eq.\eqref{eq:36}, using the values of $\Psi$ at the known grid points W, E, and S, the value of $\Psi$ at the point N can be obtained recursively.

~~The initial condition is introduced by means of a perturbation in the form of a Gaussian pulse, which is imposed on the two null surfaces   and  . Its specific mathematical expression is given by:
\begin{equation}
\Psi(u = u_0, v) = \exp\left( \frac{-(v - v_c)^2}{2\sigma^2} \right),\quad \Psi(u, v = v_0) = 0
	\label{eq:37}
\end{equation}
~~Herein, the parameter $v_c$ is used to control the characteristics of the Gaussian wave packet.

~~To extract the frequency information of QNMs from the time-domain signals obtained through numerical simulation, we perform an analysis based on the Prony method\cite{Chowdhury:2020rfj,Konoplya:2011qq}. This method extracts the frequencies and damping rates by fitting the signal as a superposition of damped exponential functions. Finally, from the time-domain evolution data, the Prony method is utilized to obtain the QNMs frequencies, and the approximate expression is given by:
\begin{equation}
	\Psi(t) \approx \sum_{k=1}^{p} C_k e^{-i\omega_k t}
	\label{eq:38}
\end{equation}

\section{Numerical Calculation of Black Hole Quasinormal Modes in the KR-PFDM Spacetime Background}
\label{sec:4}

\begin{figure*}[htbp]
	\centering
	\includegraphics[width=0.3\textwidth]{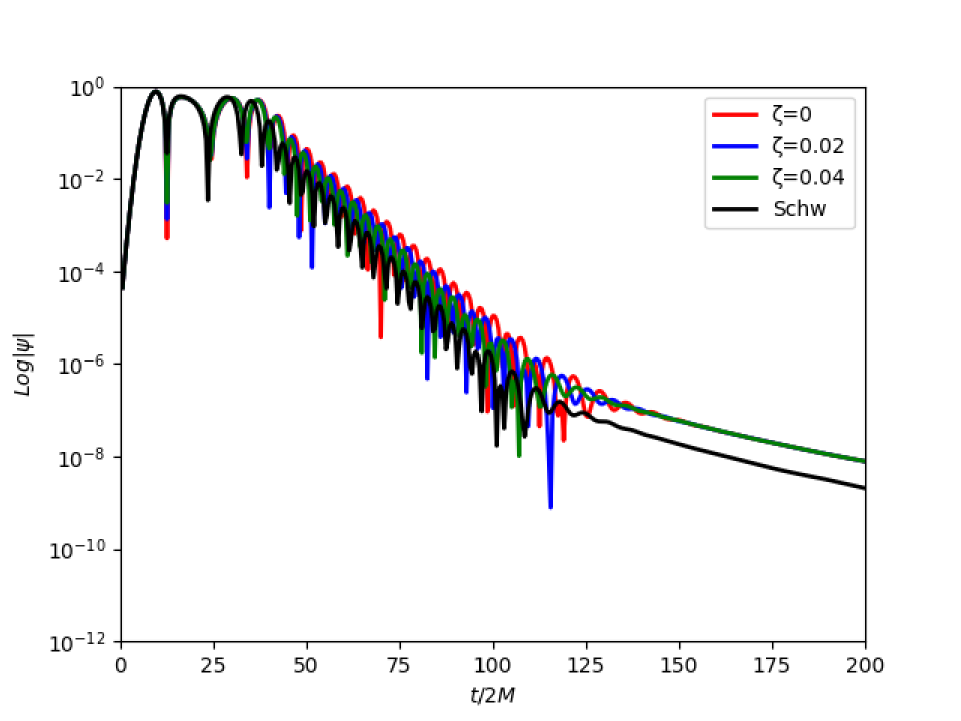} 
	\hfill
	\includegraphics[width=0.3\textwidth]{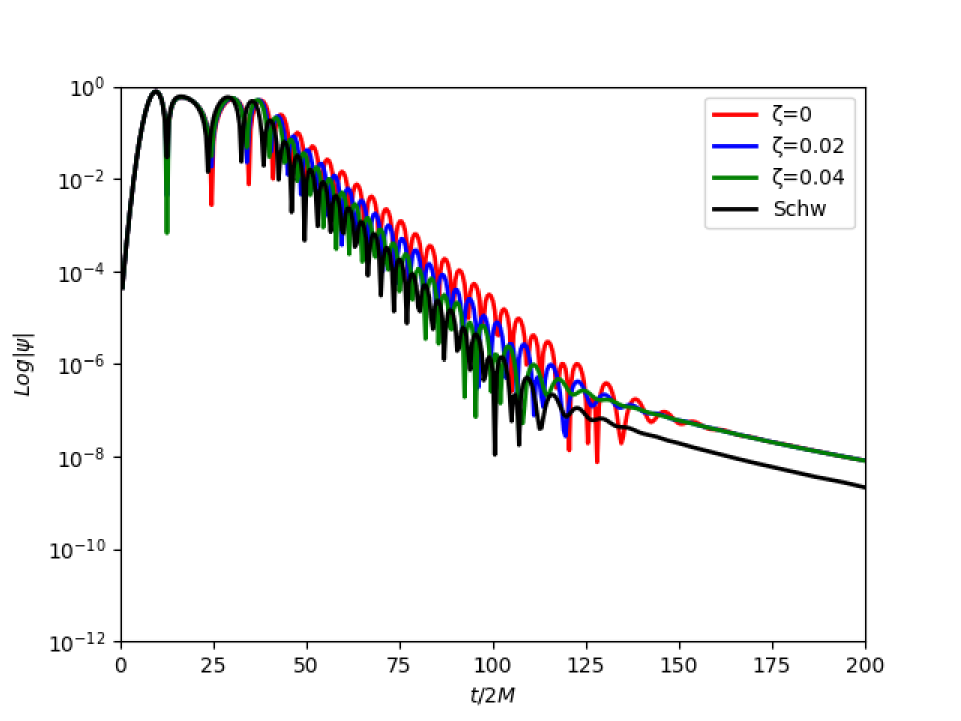} 
	\hfill
	\includegraphics[width=0.3\textwidth]{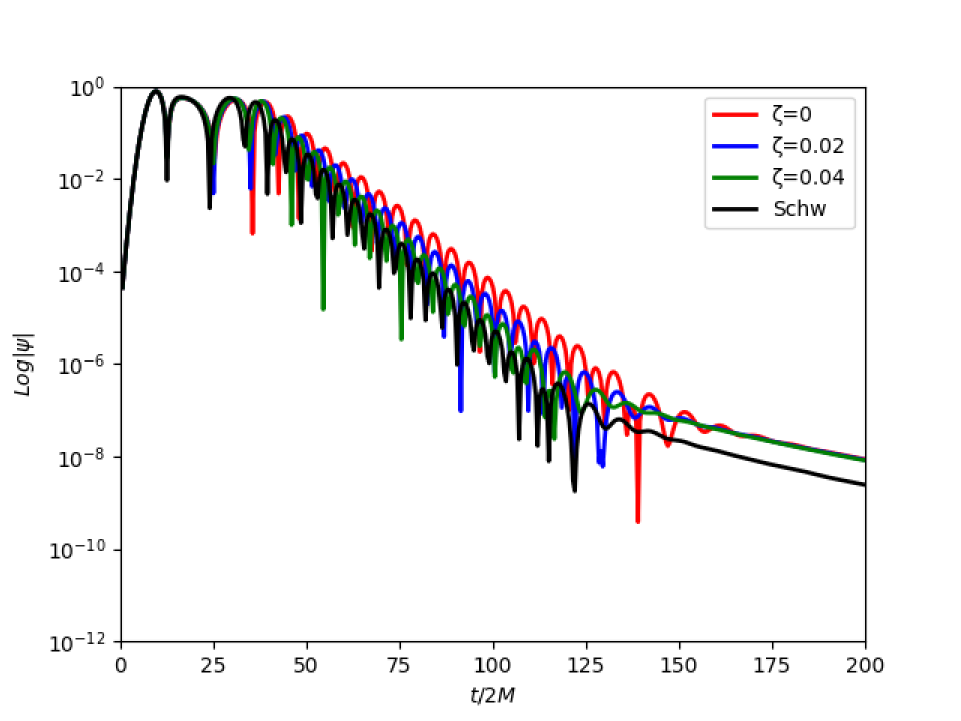} 
	\caption{Shows the time-domain evolution diagrams of QNMs under scalar field (left), electromagnetic field (middle), and gravitational field (right) perturbations for different values of $\zeta$ with fixed parameters $\tau=-0.1$, $M=1/2$, and $l=2$.}
	\label{fig:6}
\end{figure*}

\begin{figure*}[htbp]
	\centering
	\includegraphics[width=0.3\textwidth]{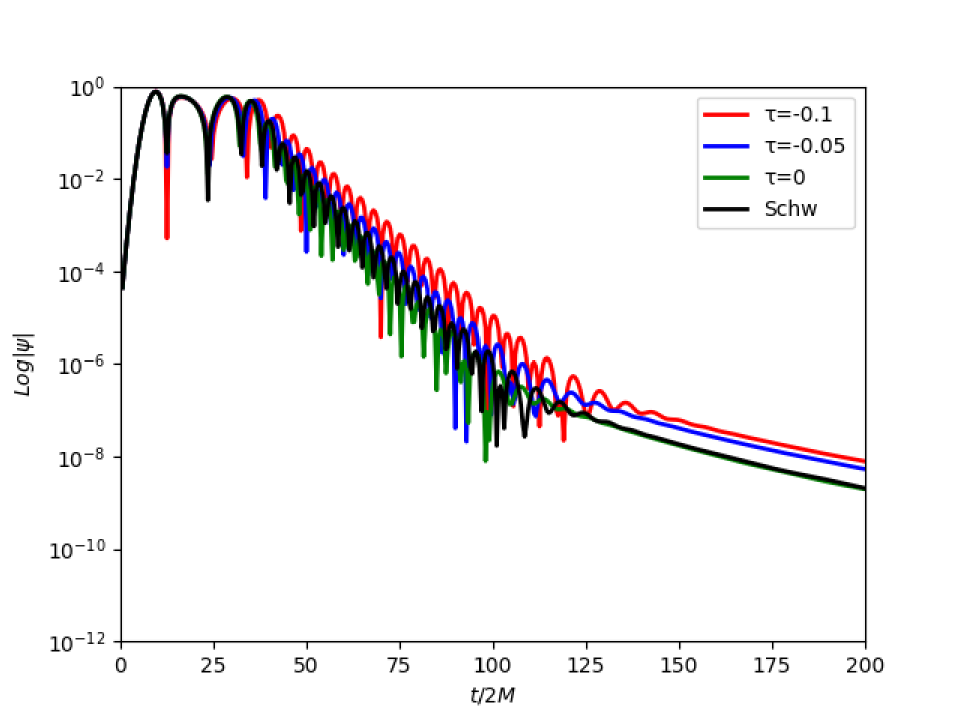} 
	\hfill
	\includegraphics[width=0.3\textwidth]{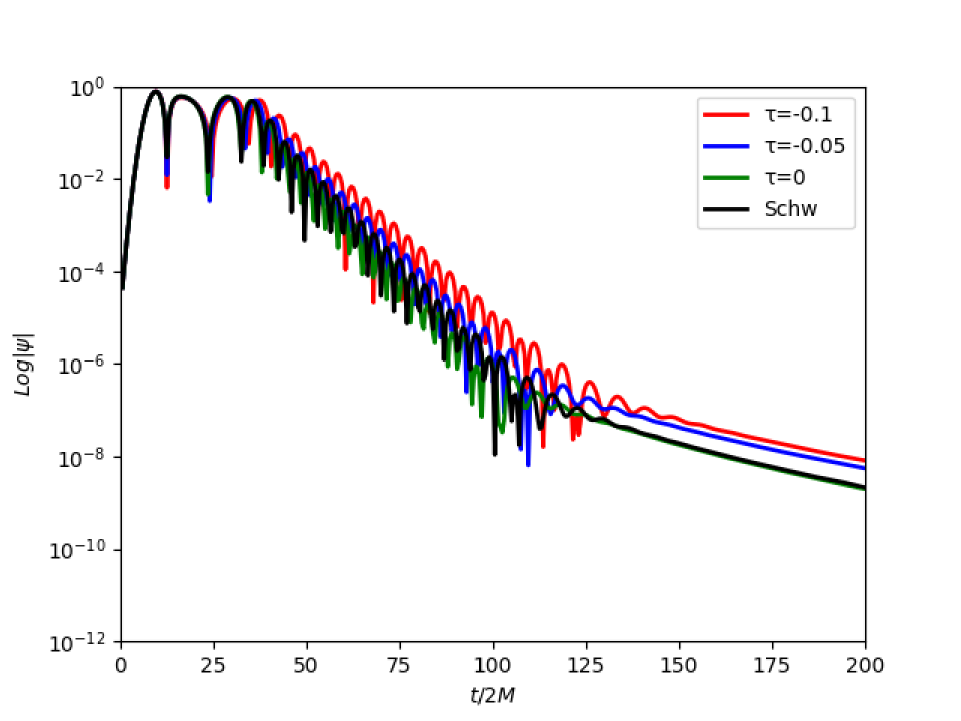} 
	\hfill
	\includegraphics[width=0.3\textwidth]{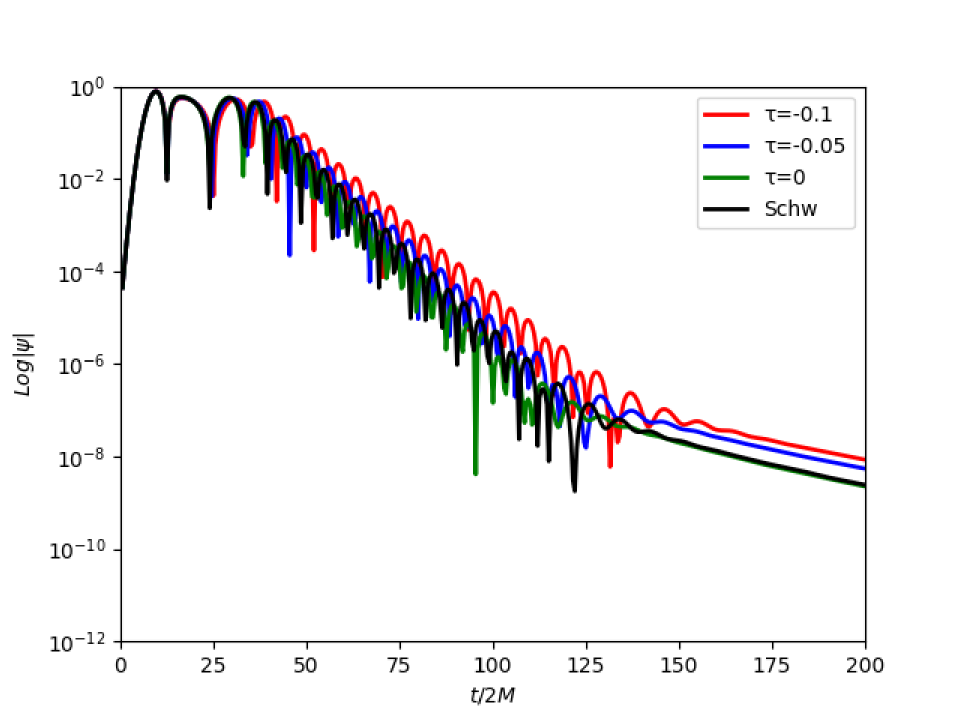} 
	\caption{Shows the time-domain evolution diagrams of QNMs under scalar field (left), electromagnetic field (middle), and gravitational field (right) perturbations for different values of $\tau$ with fixed parameters $\zeta=0.01$, $M=1/2$, and $l=2$.}
	\label{fig:7}
\end{figure*}

~~On the basis of the analysis conducted earlier, this section will further investigate the numerical calculation of black hole QNMs, focusing on the time-domain integration method and the WKB approximation method. Based on the numerical solution of the wave Eq.\eqref{eq:35}, we obtain the ringdown signal evolution waveforms corresponding to different parameter combinations. Fig.\ref{fig:6} and Fig.\ref{fig:7} respectively illustrate the dynamical evolution process of a Gaussian initial pulse in this spacetime.

~~The right, middle, and left panels correspond to the time-domain evolution curves of QNM for axial gravitational, electromagnetic, and scalar field perturbations, respectively. Under the conditions $\tau=0$ and $\zeta=0$, the system degenerates into the standard Schwarzschild black hole, and the decay behavior of the time-domain wave packet aligns with what classical gravity theory predicts. From Fig.\ref{fig:6}, it can be seen that in the early oscillatory stage, the increase in the parameter $\zeta$ has no significant effect on the perturbation waveform. However, once the system enters the QNMs-dominated phase, an increase in $\zeta$ steepens the waveform and shortens the decay time, indicating enhanced damping in the system, i.e., an increase in the absolute value of the imaginary part of the QNMs. Fig.\ref{fig:7} reveals the influence of the LSB factor $\tau$ on the time-domain evolution of QNMs. The results show that an increase in $\tau$ also leads to a steepening trend in the decay of the time-domain evolution signal. Notably, the results obtained from the time-domain integration in this section are consistent with the theoretical analysis based on the effective potential presented earlier. Together, they reveal the distinct mechanisms of the parameters $\zeta$ and $\tau$ in modulating the damping behavior of black hole oscillations.

In this paper, we first select the QNMs of the Schwarzschild black hole as a benchmark test to verify the reliability of the numerical methods. As shown in Table.\ref{tab:1}, we adopt the 
sixth-order WKB approximation and the Prony method to calculate the QNM frequencies of the Schwarzschild black hole, and compare them with the Leaver solutions in Re.\cite{Iyer:1986nq}. It is found that both methods are in excellent agreement with the Leaver solution numerically, particularly for low-order vibrational modes. The high consistency among the results obtained from various methods fully confirms the reliability and accuracy of the numerical framework employed in this paper.

\begin{table*}[!htbp]
	\centering
	\large 
	\setlength{\heavyrulewidth}{0.12ex} 
	\setlength{\lightrulewidth}{0.08ex} 
	\setlength{\tabcolsep}{12pt} 
	\begin{tabular}{cccc}
		\toprule
		$l$ & Leavers result & Prony method & WKB method  \\
		\midrule
		$l_{\text{sc}}=0$ & $0.1105-0.1049\mathrm{i}$ & $0.107711-0.1138230\mathrm{i}$ & $0.110464-0.1008190\mathrm{i}$ \\
		$l_{\text{sc}}=1$ & $0.2929-0.0977\mathrm{i}$ & $0.293554-0.0960819\mathrm{i}$ & $0.292910-0.0977616\mathrm{i}$ \\
		$l_{\text{sc}}=2$ & $0.4836-0.0968\mathrm{i}$ & $0.484244-0.0966943\mathrm{i}$ & $0.483642-0.0967661\mathrm{i}$ \\
		$l_{\text{em}}=1$ & $0.2483-0.0925\mathrm{i}$ & $0.249379-0.0918587\mathrm{i}$ & $0.248191-0.0926370\mathrm{i}$ \\
		$l_{\text{em}}=2$ & $0.4576-0.0950\mathrm{i}$ & $0.457938-0.0949696\mathrm{i}$ & $0.457539-0.0950112\mathrm{i}$ \\
		$l_{\text{gr}}=2$ & $0.3737-0.0889\mathrm{i}$ & $0.373857-0.0884667\mathrm{i}$ & $0.377945-0.1135530\mathrm{i}$ \\
		\bottomrule
	\end{tabular}
	\caption{Based on the Prony method, the sixth-order WKB approximation, and Leaver's results in the literature, this paper presents a comparative analysis of the fundamental QNM frequencies for Schwarzschild black holes under scalar, electromagnetic, and gravitational field perturbations.}
	\label{tab:1}
\end{table*}
\begin{table*}[!htbp]
	\centering
	\large 
	\setlength{\heavyrulewidth}{0.12ex} 
	\setlength{\lightrulewidth}{0.08ex} 
	\setlength{\tabcolsep}{12pt} 
	\begin{tabular}{ccccc}
		\toprule
	    Parameter $\zeta$ & WKB & Prony & $\omega_R$ (\%) & $\omega_I$ (\%) \\
		\midrule
		0     & $0.836517 - 0.159865\mathrm{i}$ & $0.838099 - 0.158847\mathrm{i}$ & 0.1888 & 0.6409 \\
		0.01  & $0.876896 - 0.168381\mathrm{i}$ & $0.875437 - 0.167514\mathrm{i}$ & 0.1667 & 0.5176 \\
		0.02  & $0.908159 - 0.175261\mathrm{i}$ & $0.907170 - 0.175323\mathrm{i}$ & 0.1090 & 0.0354 \\
		0.03  & $0.936216 - 0.181619\mathrm{i}$ & $0.936434 - 0.181436\mathrm{i}$ & 0.0233 & 0.1009 \\
		0.04  & $0.962178 - 0.187659\mathrm{i}$ & $0.963229 - 0.187143\mathrm{i}$ & 0.1091 & 0.2757 \\
		\bottomrule
	\end{tabular}
	\caption{Employing the Prony method and the sixth-order WKB approximation, the QNM frequencies corresponding to scalar field perturbations for different values of $\zeta$ are computed, given the parameter choices $M=1/2$, $l=2$, and $\tau=0.1$.}
	\label{tab:2}
\end{table*}
\begin{table*}[!htbp]
	\centering
	\large 
	\setlength{\heavyrulewidth}{0.12ex} 
	\setlength{\lightrulewidth}{0.08ex} 
	\setlength{\tabcolsep}{12pt} 
	\begin{tabular}{ccccc}
		\toprule
		Parameter $\zeta$ & WKB & Prony & $\omega_R$ (\%) & $\omega_I$ (\%) \\
		\midrule
		0     & $0.795446 - 0.157225\mathrm{i}$ & $0.796964 - 0.156255\mathrm{i}$ & 0.1905 & 0.6208 \\
		0.01  & $0.833571 - 0.165565\mathrm{i}$ & $0.832547 - 0.164810\mathrm{i}$ & 0.1230 & 0.4581 \\
		0.02  & $0.862991 - 0.172291\mathrm{i}$ & $0.862118 - 0.171549\mathrm{i}$ & 0.1013 & 0.4325 \\
		0.03  & $0.889329 - 0.178498\mathrm{i}$ & $0.888347 - 0.178480\mathrm{i}$ & 0.1105 & 0.0101 \\
		0.04  & $0.913645 - 0.184389\mathrm{i}$ & $0.913929 - 0.184111\mathrm{i}$ & 0.0311 & 0.1501 \\
		\bottomrule
	\end{tabular}
	\caption{Employing the Prony method and the sixth-order WKB approximation, the QNM frequencies corresponding to electromagnetic field perturbations for different values of $\zeta$ are computed, given the parameter choices $M=1/2$, $l=2$, and $\tau=0.1$.}
	\label{tab:3}
\end{table*}
\begin{table*}[!htbp]
	\centering
	\large 
	\setlength{\heavyrulewidth}{0.12ex} 
	\setlength{\lightrulewidth}{0.08ex} 
	\setlength{\tabcolsep}{12pt} 
	\begin{tabular}{ccccc}
		\toprule
		Parameter $\zeta$ & WKB & Prony & $\omega_R$ (\%) & $\omega_I$ (\%) \\
		\midrule
		0     & $0.663933 - 0.148069\mathrm{i}$ & $0.664414 - 0.143789\mathrm{i}$ & 0.0724 & 2.9766 \\
		0.01  & $0.694872 - 0.155835\mathrm{i}$ & $0.695847 - 0.155166\mathrm{i}$ & 0.1401 & 0.4312 \\
		0.02  & $0.718424 - 0.162068\mathrm{i}$ & $0.719538 - 0.161322\mathrm{i}$ & 0.1548 & 0.4624 \\
		0.03  & $0.739297 - 0.167804\mathrm{i}$ & $0.739093 - 0.166426\mathrm{i}$ & 0.0276 & 0.8280 \\
		0.04  & $0.758384 - 0.173231\mathrm{i}$ & $0.758426 - 0.172125\mathrm{i}$ & 0.0055 & 0.6426 \\
		\bottomrule
	\end{tabular}
	\caption{Employing the Prony method and the sixth-order WKB approximation, the QNM frequencies corresponding to axial gravitational field perturbations for different values of $\zeta$ are computed, given the parameter choices $M=1/2$, $l=2$, and $\tau=0.1$.}
	\label{tab:4}
\end{table*}
\begin{table*}[!htbp]
	\centering
	\large 
	\setlength{\heavyrulewidth}{0.12ex} 
	\setlength{\lightrulewidth}{0.08ex} 
	\setlength{\tabcolsep}{12pt} 
	\begin{tabular}{ccccc}
		\toprule
		Parameter $\tau$ & WKB & Prony & $\omega_R$ (\%) & $\omega_I$ (\%) \\
		\midrule
		-0.1  & $0.876896 - 0.168381\mathrm{i}$ & $0.875437 - 0.167514\mathrm{i}$ & 0.1667 & 0.5176 \\
		-0.05 & $0.943009 - 0.185223\mathrm{i}$ & $0.944331 - 0.184456\mathrm{i}$ & 0.1400 & 0.4158 \\
		0     & $1.01784  - 0.204719\mathrm{i}$ & $1.02066  - 0.20332 \mathrm{i}$ & 0.2763 & 0.6881 \\
		0.025 & $1.05903  - 0.21564 \mathrm{i}$ & $1.06267  - 0.213257\mathrm{i}$ & 0.3426 & 1.1174 \\
		0.05  & $1.10307  - 0.227458\mathrm{i}$ & $1.10715  - 0.225053\mathrm{i}$ & 0.3685 & 1.0686 \\
		\bottomrule
	\end{tabular}
	\caption{Employing the Prony method and the sixth-order WKB approximation, the QNM frequencies corresponding to scalar field perturbations for different values of $\tau$ are computed, given the parameter choices $M=1/2$, $l=2$, and $\zeta=0.01$.}
	\label{tab:5}
\end{table*}
\begin{table*}[!htbp]
	\centering
	\large 
	\setlength{\heavyrulewidth}{0.12ex} 
	\setlength{\lightrulewidth}{0.08ex} 
	\setlength{\tabcolsep}{12pt} 
	\begin{tabular}{ccccc}
		\toprule
		Parameter $\tau$ & WKB & Prony & $\omega_R$ (\%) & $\omega_I$ (\%) \\
		\midrule
		-0.1  & $0.833571 - 0.165565\mathrm{i}$ & $0.83521 - 0.164493\mathrm{i}$ & 0.1962 & 0.6517 \\
		-0.05 & $0.894248 - 0.181978\mathrm{i}$ & $0.894874 - 0.181279\mathrm{i}$ & 0.0700 & 0.3856 \\
		0     & $0.962637 - 0.200954\mathrm{i}$ & $0.965032 - 0.19953\mathrm{i}$ & 0.2482 & 0.7137 \\
		0.025 & $1.00016 - 0.211573\mathrm{i}$ & $1.00307 - 0.209855\mathrm{i}$ & 0.2901 & 0.8187 \\
		0.05  & $1.04018 - 0.223055\mathrm{i}$ & $1.04354 - 0.220875\mathrm{i}$ & 0.3220 & 0.9870 \\
		\bottomrule
	\end{tabular}
	\caption{Employing the Prony method and the sixth-order WKB approximation, the QNM frequencies corresponding to electromagnetic field perturbations for different values of $\tau$ are computed, given the parameter choices $M=1/2$, $l=2$, and $\zeta=0.01$.}
	\label{tab:6}
\end{table*}
\begin{table*}[!htbp]
	\centering
	\large 
	\setlength{\heavyrulewidth}{0.12ex} 
	\setlength{\lightrulewidth}{0.08ex} 
	\setlength{\tabcolsep}{12pt} 
	\begin{tabular}{ccccc}
		\toprule
		Parameter $\tau$ & WKB & Prony & $\omega_R$ (\%) & $\omega_I$ (\%) \\
		\midrule
		-0.1  & $0.694872 - 0.155835\mathrm{i}$ & $0.695831 - 0.155152\mathrm{i}$ & 0.1378 & 0.4402 \\
		-0.05 & $0.737653 - 0.170739\mathrm{i}$ & $0.738957 - 0.169945\mathrm{i}$ & 0.1765 & 0.4672 \\
		0     & $0.784718 - 0.187895\mathrm{i}$ & $0.785679 - 0.187185\mathrm{i}$ & 0.1223 & 0.3793 \\
		0.025 & $0.810019 - 0.197467\mathrm{i}$ & $0.812026 - 0.196286\mathrm{i}$ & 0.2472 & 0.6017 \\
		0.05  & $0.836592 - 0.207801\mathrm{i}$ & $0.838751 - 0.206541\mathrm{i}$ & 0.2574 & 0.6100 \\
		\bottomrule
	\end{tabular}
	\caption{Employing the Prony method and the sixth-order WKB approximation, the QNM frequencies corresponding to axial gravitational field perturbations for different values of $\tau$ are computed, given the parameter choices $M=1/2$, $l=2$, and $\zeta=0.01$.}
	\label{tab:7}
\end{table*}
~~To investigate the behavior of QNMs for black holes in the KR-PFDM spacetime background, this paper employs both the Prony method and the sixth-order WKB approximation method to numerically solve the QNM frequencies corresponding to various perturbation fields. Table\ref{tab:2} to \ref{tab:7} summarize the relevant results and systematically show the influence of the PFDM parameter $\zeta$ and the LSB factor $\tau$ on the QNM frequencies of the black hole at the center of the M87* galaxy.

~~First, under the condition of a fixed LSB factor $\tau=0.1$ , the data in Tables\ref{tab:2} to \ref{tab:4} indicate that as the PFDM parameter $\zeta$ increases, both the real portion and the absolute magnitude of the imaginary portion of the QNM frequencies exhibit a strictly monotonically increasing trend. This result suggests that $\zeta$ and $\tau$ do not independently influence black hole perturbations; instead, through the coupling terms in the KR-PFDM theory, they jointly modify the energy-momentum tensor and spacetime curvature, thereby altering the shape of the effective potential barrier and producing an equivalent "gravitational stiffening" effect. Moreover, the unique logarithmic correction term $\frac{\zeta}{r(1-\tau)} \ln\left( \frac{r}{|\zeta|} \right)$ of this theoretical model may also act to “stiffen” spacetime, further strengthening this trend.

~~Secondly, on the premise that the PFDM parameter $\zeta=0.01$ is fixed, the data in Tables\ref{tab:5} to \ref{tab:7} further reveal the influence of the LSB factor $\tau$ on the QNMs. The numerical results indicate that as the value of $\tau$ increases, both the real portion and the absolute magnitude of the imaginary portion of the corresponding QNMs frequencies for various perturbation fields also exhibit a monotonically increasing trend. This rule of variation is highly consistent with the scenario dominated by the PFDM parameter $\zeta$, suggesting that within the framework of the KR-PFDM theory, PFDM and LSB effects jointly influence the perturbation response of the black hole through coupling terms in the theory. The physical mechanism can be traced back to the profound modification of spacetime geometry by the LSB effect: as the value of $\tau$ increases, the value of the logarithmic correction term  $\frac{\zeta}{r(1-\tau)} \ln\left( \frac{r}{|\zeta|} \right)$ of the metric function in the KR-PFDM theory increases accordingly, leading to an overall elevation of the effective potential barrier (as illustrated in Fig.\ref{fig:5}). A higher potential barrier implies a stronger ability of spacetime to confine external perturbations, simultaneously resulting in faster oscillation frequencies of the QNMs (increase in the real part) and enhanced efficiency of energy dissipation (rise in the magnitude of the imaginary portion).

~~Notably, the aforementioned evolution rule stands in stark contrast to the typical characteristics predicted by most traditional dark matter models, where dark matter leads to a decrease in the oscillation frequency and damping rate of QNMs. This contrast highlights the unique spacetime dynamical properties induced by the coupling between PFDM and LSB effects within the framework of the KR-PFDM theory. It also provides a new theoretical basis for using gravitational wave ringdown signals to discriminate between different spacetime symmetry-breaking effects and dark matter models in future observations.

\section{Summary}
\label{sec:5}

In this study, within the theoretical framework where spontaneous LSB induced by KR field coexists with PFDM, the perturbative dynamics of static spherically symmetric black holes are systematically investigated. To ensure that the values of the LSB factor $\tau$ and the PFDM parameter $\zeta$ fall within their theoretical ranges, we constrain these parameters using the EHT observational data on the shadow of the M87* black hole. The numerical calculation results yield an interesting conclusion: the synergistic effect of $\tau$ and $\zeta$ does not exhibit the commonly expected spacetime "softening" behavior, but rather manifests as a "stiffening" effect (i.e., accelerated oscillations and enhanced dissipation).This result reveals a potentially unique coupling mechanism between the KR field and dark matter in the perturbative dynamics of black holes, provides a concrete example for understanding the ringdown behavior of black holes in the modified gravity framework, and contributes to a deeper comprehension of black hole dynamics under modified gravity theories.

\section*{Acknowledgments}
This research partly supported by the
National Natural Science Foundation of China (Grant No. 12265007).

\newpage


\bibliographystyle{unsrt}  
\bibliography{ref1}
\bibliographystyle{apsrev4-1}

\end{document}